\documentclass[11pt]{article} 
\usepackage[utf8]{inputenc}
\usepackage{multirow}
\usepackage{amsfonts}
\usepackage{amsmath}
\usepackage{amssymb}
\usepackage{amsthm}
\usepackage[export]{adjustbox}
\usepackage[english]{babel}
\usepackage{tikz}
\usetikzlibrary{3d}
\usetikzlibrary{shapes.misc}
\usetikzlibrary{decorations.markings,decorations.pathmorphing}
\usetikzlibrary{arrows}
\tikzset{cross/.style={cross out, draw=black, minimum size=2*(#1-\pgflinewidth), inner sep=0pt, outer sep=0pt}, cross/.default={1pt}}
\usetikzlibrary{decorations.pathmorphing}

\makeatletter

\usepackage{comment}
\usepackage{float}

\usepackage{caption}
\usepackage{xcolor}
\definecolor{dgreen}{rgb}{0,0.5,0}
\definecolor{darkblue}{rgb}{0,0,0.6}
\definecolor{lblue}{rgb}{0.5,0.5,1.0}
\definecolor{purple}{rgb}{0.4,.2,0.7}
\definecolor{ggray}{rgb}{0.85,.85,0.85}

\usepackage[margin = 2.5cm]{geometry}
\usepackage{graphicx}
\usepackage[hyperfootnotes = false, colorlinks = true, linkcolor = blue, citecolor = purple]{hyperref}
\usepackage{subcaption}

\def\la{\label}

\newcommand{\bes}{\begin{equation} \begin{split} }	
	\newcommand{\ees}{\end{split} \end{equation} }
	
	\renewcommand{\i}{\mathrm{i}}

	\newcommand{\pd}{\partial}

	\newcommand{\hf}{\tfrac{1}{2}}
	
	\usepackage{physics}

\usepackage{amsfonts}
\usepackage{amsmath}
\usepackage{amssymb}
\usepackage{braket}
\usepackage{cite}
\usepackage{enumitem}
\usepackage{float}
\usepackage{graphicx}
\usepackage{mathrsfs}
\usepackage{setspace}
\usepackage{slashed}
\usepackage{physics}
\usepackage{subcaption}
\usepackage{titlesec}
\usepackage{comment}

\def\hf{\tfrac{1}{2}}
\def\la{\label}
\def\nref#1{(\ref{#1})}

\usepackage[margin = 2.5cm]{geometry}
\usepackage{color}
\definecolor{darkblue}{rgb}{0,0,0.6}
\definecolor{purple}{rgb}{0.4,.2,0.7}
\definecolor{darkgreen}{rgb}{0,0.5,0}

\renewcommand{\i}{\mathrm{i}}
\renewcommand{\d}{\mathrm{d}}
\newcommand{\ads}{\mathrm{AdS}}

\begin{document}

\thispagestyle{empty}
\begin{center}
    ~\vspace{5mm}

     {\LARGE \bf Giant gravitons in D$p$-brane holography}

   \vspace{0.5in}
     
   {\bf  Gauri Batra$^a$ \& Henry W. Lin$^{a,b}$ }

    \vspace{0.5in}

   ~
   \\
   {$^a$ Stanford Institute for Theoretical Physics, Stanford University, Stanford, CA 94305, USA}  \\
   {$^b$ Jadwin Hall, Princeton University, Princeton, NJ 08540, USA}
                
    \vspace{0.5in}

    \vspace{0.5in}
    
\end{center}

\vspace{0.5in}

\begin{abstract}

We consider half BPS operators in maximally supersymmetric Yang Mills (SYM) in $p+1$ dimensions. These operators satisfy trace relations that are identical to those discussed in the $p=3$ case ($\mathcal{N} = 4$ SYM). 
Nevertheless, the bulk explanation of these trace relations must differ from the $p = 3$ case as their holographic duals are not AdS spacetimes.
We identify giant graviton solutions in the dual holographic backgrounds for $-1 \le p \le 4$. 
In the 't Hooft limit, these giants are D$(6-p)$-branes that wrap a $S_{6-p} \subset S_{8-p}$. We also follow the giants into the strong coupling region where they become other branes.
Despite propagating in a non-AdS geometry, we find that the branes ``feel'' like they are in AdS. This is closely related to the emergent scaling symmetry present in these boundary theories.

\end{abstract}

\vspace{1in}

\pagebreak

\setcounter{tocdepth}{3}

\tableofcontents

\section{Introduction}

Holographic theories with known Lagrangian descriptions typically involve matrix-valued fields. %
Matrices satisfy trace relations at finite $N$. 
These trace relations kick in when one considers composite operators made out of $\sim N$ fields; they are therefore associated with D-branes in the bulk \cite{McGreevy:2000cw, Grisaru:2000zn, Hashimoto:2000zp}.
Roughly speaking, as one dials up the angular momentum of a graviton on the internal sphere, it blows up into a wrapped D-brane  (and is therefore ``giant'').\footnote{Recently, \cite{Chang:2022mjp} proposed that trace relations for $\frac{1}{16}$-BPS operators play an important role in understanding black hole microstates; see also \cite{Choi:2022caq, Choi:2023znd, Budzik:2023vtr, Choi:2025lck}.}

Much of the existing literature on giant gravitons has concentrated on conformal field theories, such as $\mathcal{N}= 4$ super Yang-Mills (SYM) \cite{McGreevy:2000cw,Grisaru:2000zn, Hashimoto:2000zp,Balasubramanian:2001nh,Corley:2001zk,gaiotto2021giantgravitonexpansion,imamura2021finitensuperconformalindexadscft,murthy2023unitarymatrixmodelsfree}.
But trace relations have little to do with conformal symmetry and therefore we would like to understand their bulk interpretation in non-conformal examples of holography. Perhaps the simplest example of non-conformal holography is to consider SYM in spacetime dimensions $p+1$ with maximal supersymmetry. Since SYM can be viewed as the low-energy description of a stack of $N$ D$p$-branes, the holographic dual is well known \cite{Itzhaki:1998dd}. %
As we review in Section \ref{sec:review}, the 1/2 BPS operators in these SYM theories are multi-traces of a complex matrix. These operators satisfy trace relations identical to those often discussed in the context of $\mathcal{N} = 4$ SYM ($p=3)$. Nevertheless, the bulk explanation of these trace relations must be somewhat different for $p \ne 3$, since for $p \ne 3$ SYM is not a conformal field theory and its holographic dual is not AdS.

We identify giant graviton solutions in the black brane gravity solutions dual to SYM for $1 \le p \le 4$. They are given by D$(6-p)$ branes that wrap an $S_{6-p}\subset S_{8-p}$ in the spacetime and carry SO($9-p$) angular momentum on the sphere. There exists a maximal giant graviton that wraps a maximal $S_{6-p}$ and saturates the inequality $J \leq N$; this is the bulk explanation of the trace relations. This can be further sharpened by quantizing small fluctuations around the maximal giant \cite{Lee:2024hef}; we discuss this further in Appendix \ref{app:quantum}. An interesting feature of the bulk solutions is that the giants ``feel'' like they are living in the AdS$_{d+1}$ black brane. This means for instance that in the radial/time directions the giants do not propagate on geodesics of the true geometry but rather propagate on geodesics of the AdS$_{d+1}$ black brane. This is closing related to the scaling symmetry discussed in \cite{biggs2023scaling}; see also \cite{Boonstra:1998mp,Sekino:1999av,Dong:2012se}.

In non-conformal holography, the dilaton grows (either towards the IR for $p<3$ or towards the UV for $p>3$). We follow the giants into these strongly coupled regions in section \ref{sec:discuss}. After applying dualities the giants become other branes. For some values of $p$, we show that after applying the relevant dualities the branes continue to ``feel'' like they are in AdS.

Part of our motivation to understand these giant gravitons more generally is that there has been recent progress in boundary techniques for analyzing the low-dimensional cases $p=0$ \cite{Lin:2023owt, Cho:2024kxn, Lin:2024vvg, Pateloudis:2022ijr} and $p=-1$ theories \cite{Ishibashi:1996xs}, see \cite{Hartnoll:2024csr, Komatsu:2024bop, Komatsu:2024ydh}. 
Using boundary techniques it may be possible to learn properties of these giants and/or the bulk spacetimes that they probe.

\section{Review of D\texorpdfstring{$p$}{p}-brane holography \label{sec:review}}
Let us review the gravity dual of SYM in $p+1$ spacetime dimensions. We view the SYM theory as arising from a stack of $N$ D$p$-branes. From the bulk perspective, we consider the black brane solutions and then take the decoupling limit \cite{Itzhaki:1998dd, Maldacena:1997re}\footnote{Let $u$ be a boundary timescale/lengthscale. In the decoupling limit, we hold $\tau \sim  u /(g^2_\text{YM})^{1/{(p-3)}}$ fixed while taking $u/\ell_s \to 0$. This suppresses higher derivative corrections to the SYM action, as well as couplings between the D$p$ branes and the bulk. Note that for $p<3$ we are taking the asymptotic string coupling $g_s \to 0$ whereas for $p>3$ we are taking $g_s \to \infty$.}. We use conventions where the relevant part of the Type II supergravity action is
\begin{align}\label{actionII}
   I =  \frac{1}{(2\pi)^7 (\alpha')^4} \int \d^{10} x \sqrt{g} \left[e^{-2 \phi} (R+4 (\nabla \phi)^2) -\frac{1}{2 (p+2)!} F_{p+2}^2\right]
\end{align}

The string-frame solutions (for $p\ne 3$) are:
\begin{align}
    \la{gzz}  \frac{ \d s^2}{\alpha'} &= \left(\frac{z}{R_\ads }\right)^{\frac{3-p}{5-p}} \left[R_\ads^2 \left(\frac{h(z)\, \d \tau^2+ h^{-1}(z) \d z^2 +\d \vec{x}_p^2}{z^2}\right)+\d \Omega_{8-p}^2\right], \\
    h&=1-\frac{z^d}{z_0^d}, \quad  d = 1+\frac{9-p}{5-p}, \quad 
    R_\ads = \frac{2}{5-p}, \quad  d_p =2^{7-2p}\pi^{\frac{9-3p}{2}}\Gamma\left( \hf (7-p) \right), \\
    e^{-2\phi} &=  (d_p (2\pi)^{p-2}N)^{2} \left(\frac{z}{R_\ads}\right)^{\frac{7-p}{5-p} (p-3)} \label{phi_sol},\\
    A_{0 \cdots p } &=  (\sqrt{\alpha'})^{p+1} d_p (2\pi)^{p-2}N\left(\frac{z}{R_\ads}\right)^{-2\frac{7-p}{5-p}}. 
\end{align}
Here, the string coupling is related to the Yang-Mills coupling constant via $g^2_\mathrm{YM} = (2\pi)^{p-2} g_s \ell_s^{p-3}$ where $\alpha ' = \ell_s^2$. Note that the 't Hooft coupling $g^2_\text{YM} N$ is dimensionful and has units of $\ell_s^{p-3}$. At finite temperature $1/\beta$, the 't Hooft limit is $g^2_\text{YM} N / \beta^{p-3} = \text{fixed}$.
We have chosen coordinates such that the metric is obviously conformal to the AdS black brane solution $\times$ sphere\footnote{At zero temperature, the metric is conformal to $\ads_{p+2} \times S_{8-p}$. The $z$ coordinate is related to the more familiar radial coordinate in \cite{Itzhaki:1998dd} via $z \propto r^{(p-5)/2}$.}.

It is instructive to consider the action associated with small fluctuations of the dilaton, e.g., $\phi = \phi_\text{s} + \varphi$, where the classical solution $\phi_\text{s}$ is given by \nref{phi_sol}. (The treatment below is somewhat heuristic\footnote{Among other concerns, we have not argued that $\varphi$ does not mix with other modes.}; for a detailed treatment, see \cite{Sekino:1999av} and also \cite{biggs2023scaling}.) Then expanding to quadratic order in $\varphi$, \eqref{actionII} gives %
\begin{align}\label{dil1}
    I &= \frac{1}{(2\pi)^7 (\alpha')^4} \int \d^{10} x \sqrt{g} e^{-2\phi_\text{s}}\left[2\varphi^2R+4 (\nabla \varphi)^2-16 \varphi  \nabla_\mu \phi_\text{s} \nabla^\mu \varphi + 8 \varphi^2 (\nabla \phi_\text{s})^2 \right]\\
    &= \frac{1}{(2\pi)^7 (\alpha')^4} \int \d^{10} x \sqrt{g} e^{-2\phi_\text{s}}\left[2\varphi^2R+4 (\nabla \varphi)^2-4 \varphi^2 e^{2\phi_\text{s}} \nabla^2 (e^{-2\phi_\text{s}}) + 8 \varphi^2 (\nabla \phi_\text{s})^2 \right] \label{dil2}\\
    &= \frac{1}{(2\pi)^7 (\alpha')^4} \int \d^{10} x \sqrt{g} e^{-2\phi_\text{s}}\left[2\varphi^2 (R+4 \nabla^2 \phi_\text{s} - 4 (\nabla \phi_\text{s})^2 )+4 (\nabla \varphi)^2 \right]\\
    &= \frac{4}{(2\pi)^7 (\alpha')^4} \int \d^{10} x \sqrt{g} e^{-2\phi_\text{s}}  (\nabla \varphi)^2.\label{probefield}
\end{align}
In writing the first line \eqref{dil1}, we have used the equations of motion to remove terms linear in $\varphi$. In  \eqref{dil2} we have dropped a total derivative. In the last line we have used the equations of motion $R+4 \nabla^2 \phi_\text{s} - 4 (\nabla \phi_\text{s})^2 = 0$ to arrive at the simple action \eqref{probefield}.
Decomposing $\varphi$ into eigenfunctions of the scalar Laplacian on the sphere, and using that the eigenvalues are $\nabla_{S^n} = - k(k+n-1)$, we have:
\begin{align}\label{action}
    I&\propto \int \d^{8-p} \Omega \,  \d z \, \d \tau \,\d^p \vec{x} \,  z^{1-d}  \left[ R^2_\ads \left( h (\pd_z \varphi)^2 +  \frac{(\pd_\tau \varphi)^2}{h} +(\vec{\nabla} \varphi)^2 \right) + k (k + 7-p) \varphi^2 \right] \\
    &=  \int \d^{8-p} \Omega \, \d^{d-1} \vec{x} \, \d z \, \d \tau\, \sqrt{g_\text{AdS}}  \left[ (\nabla_{\ads} \varphi)^2 +m_k^2 \varphi^2\right]  %
\end{align}
We see that these modes behave like massive fields in the AdS$_{d+1}$ black brane \cite{Kanitscheider:2009as, biggs2023scaling}, with an effective mass $m_k^2 = k(k+7-p)$. Note that we have introduced an additional $d-(p+1)$ spatial dimensions to account for the factors of $z$. (We implicitly assume that these extra dimensions are toroidal $T^{d-(p+1)}$ and that all fields are homogeneous on $T^{d-(p+1)}$, see \cite{biggs2023scaling}.)
One can then compute the scaling dimensions of these fields using the usual relation 
\begin{align}
    m^2 R^2_\ads = \Delta ( \Delta- d) \implies    \Delta = R_\ads (k + 2) + 2 \label{dim1}
\end{align}
Note here the scaling dimension is defined so that (in the zero-temperature limit) %
\begin{align} \label{2pt}
    \ev{\mathcal{O_\phi}(x) \mathcal{O_\phi}(0)} \sim \frac{1}{|x|^{2(\Delta - d-p-1)}}.
\end{align}
Here the additional factor of $d-(p+1)$ arises since $\mathcal{O}$ is a zero mode in the ``extra'' AdS dimensions.

This gravity analysis shows there should be boundary operators that have conformal 2-pt functions \cite{Sekino:1999av, Kanitscheider:2008kd, Kanitscheider:2009as}. We now review a schematic argument to determine the form of the operator in the boundary theory (for a review in the AdS context, see section 4 of \cite{Klebanov:2000me}). Before taking the decoupling limit, we consider a background with some (weak-field) dilaton waves incident on the D$p$-branes. Then in the decoupling limit, we are left with a perturbation of flat space with some dilaton waves $\otimes$ the throat-like region given by \eqref{gzz} but with a non-trivial perturbation to the dilaton profile. 
On the boundary side, the DBI action contains a coupling $e^{-\phi({x,X})} F^2$ where $F^2$ is the Yang-Mills field strength and $X^I$ are the transverse coordinates of the D$p$-brane.
These scalars transform under the global $R$-symmetry of the SYM as SO($9-p$) fundamentals. The coupling to the dilaton wave induces a perturbation which leads to a deformation of the worldvolume action\cite{Klebanov:1997kc, Klebanov:1999xv} by terms schematically of the form 
\begin{align}
    S_\text{SYM} \to   S_\text{SYM} + \mathcal{N} \sum_j \frac{1}{k!}\int \d^{p+1} x\,  \pd_{I_1} \cdots \pd_{I_k} \varphi \Tr (F^2_{\mu \nu} X^{(I_1} \cdots X^{I_k)}).
\end{align}
Here the indices $I_1\cdots I_k$ are symmetrized with traces removed\footnote{The indices should be symmetrized since the order of the derivatives acting on $\phi$ does not matter.}. The precise form of the operator (including the fermionic terms) can be obtained by acting with 4 supercharges on the operator $\Tr X^{(I_1} \cdots X^{I_{k+2})}$. Since acting with a supercharge changes the dimension of an operator by $1/2$, together with \eqref{dim1}, this implies that the symmetric traceless operator $\Tr X^{(I_1} \cdots X^{I_{k})}$ has dimension
\begin{align}
\label{eq:dim}
    \Delta_{\mathcal{O}} = R_\ads k, \quad  \mathcal{O}_k = \Tr X^{(I_1} \cdots X^{I_{k})}.
\end{align}
Let's recall the supersymmetric properties of the operator $\mathcal{O}_k$. In the above $\mathcal{O}_k$ denotes any one of a number of operators that spans the symmetric traceless irrep of SO($9-p$). To analyze the action of the supercharge on this operator, it is convenient to pick one particular element of the irrep. We define the complex matrix 
\begin{align}
    \mathcal{Z}=X^{1} + \i X^{2},\quad O_\text{BPS} =  \Tr \mathcal{Z}^k.
\end{align}
Clearly $\mathcal{Z}^{\otimes k}$ transforms under SO($9-p$) as a symmetric tensor of rank $k$; it is also traceless since $\mathcal{Z} = \vec{y} \cdot \vec{X}$ and $\vec{y}$ is a null vector $\vec{y} \cdot \vec{y} = 0$.
Note that $[Q_\alpha, \mathcal{Z}] = (\gamma^1 + \i \gamma^2)_{\alpha \beta} \psi^\beta$. The matrix $(\gamma^1 + \i \gamma^2)$ has a $\mathcal{N}/2$ dimensional null space. (Here $\mathcal{N}$ is the dimension of the spinor irrep of SO($9-p$), e.g., $\mathcal{N}= 16$ for the D0 brane theory and famously $\mathcal{N} = 4$ for $p=3$.)
So any operator made out of $\mathcal{Z}$ is 1/2-BPS. In summary, the 1/2-BPS operators in general SYM theories are in the same super-multiplet as dilaton modes on the sphere; similar to the case of $\mathcal{N} =4$ SYM, the 2-pt function of such operators satisfies \eqref{2pt} with dimension \eqref{eq:dim}.%

We discussed single trace $\hf$-BPS operators above, but more generally a $\hf$-BPS operators is a multi-trace of $\mathcal{Z}$: %
\begin{align} \label{multi_trace}
\mathcal{O}_\text{BPS} = \Tr \mathcal{Z}^{k_1} \Tr \mathcal{Z}^{k_2} \cdots \Tr \mathcal{Z}^{k_n}, \quad J = \sum_{i=1}^n k_i.
\end{align}
At finite $N$, these operators satisfy trace relations, which have been extensively discussed in $\mathcal{N} = 4$ SYM. A more convenient basis for multi-traces is one where different operators are orthogonal. This is the Schur basis, labeled by an irrep $R$ of SU$(N)$ \cite{Corley:2001zk}:
\begin{align}
    \mathcal{O}_R = \chi_R(\mathcal{Z}). \label{eq:char}
\end{align}
Here $\chi_R$ is a character for the irrep $R$; irreps are labeled by Young diagrams with $J$ boxes. For anti-symmetric tensors (a tall column), the maximum height of the column\footnote{For $J=N$ we may write \cite{Balasubramanian:2001nh} $\mathcal{O}_R = \det \mathcal{Z}$. More generally the column states are sub-determinants \cite{Balasubramanian:2001nh}.} is $J \le N$. This fact can be repackaged by defining a sum over single column operators:
\begin{align}
    \sum_\text{\text{single column operators}} q^J = \frac{1}{1-q} - \frac{q^{N+1}}{1-q}. \label{eq:sum}
\end{align}
This is the simplest example of the giant graviton expansion \cite{gaiotto2021giantgravitonexpansion,Lee:2024hef,imamura2021finitensuperconformalindexadscft, Lee:2023iil}, see also \cite{murthy2023unitarymatrixmodelsfree, Liu:2022olj,Eniceicu:2023uvd,Chen:2024cvf}. The point of this expression is that the first term is the $N=\infty$, whereas the second term (which is negative) is interpreted as a correction coming from the finite $N$ trace relations.
Since the second term is of order $\sim e^{-\# N}$, it is naturally associated to some kind of D-brane effect; recently  \cite{Lee:2024hef, Eleftheriou_2024,eleftheriou2025localizationwallcrossinggiantgraviton} showed that a quantum analysis of the maximal giant graviton spectrum (for the $p=3$ case) reproduces the second term in \eqref{eq:sum} including the minus sign. Here we are emphasizing that  the {\it boundary} analysis is essentially uniform in $p$; in Section \ref{sec:solution} and in Appendix \ref{app:quantum} we will argue that their bulk analysis generalizes to $p \ne 3$ as well.

\section{Giant graviton solutions \label{sec:solution}}
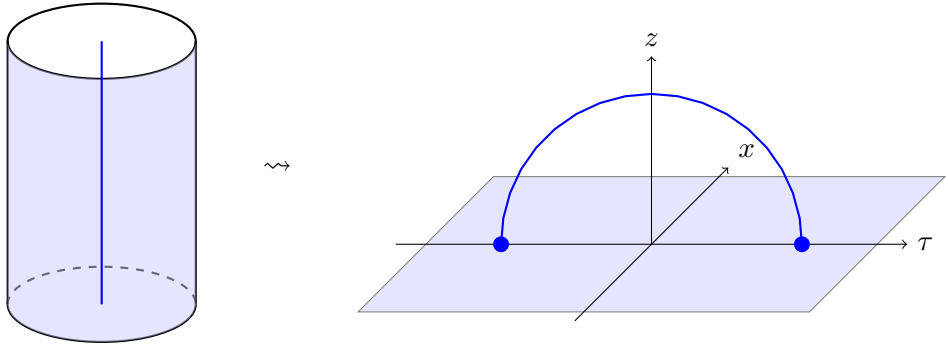
\begin{figure}[H]
    \centering
\begin{align*}
\begin{tikzpicture}[baseline={([yshift=0cm]current bounding box.center)}]
\draw[thick] (0,0) ellipse (1.25 and 0.5);
\draw[thick] (-1.25,0) -- (-1.25,-3.5);
\draw[thick] (-1.25,-3.5) arc (180:360:1.25 and 0.5);
\draw[thick,dashed] (-1.25,-3.5) arc (180:360:1.25 and -0.5);
\draw[thick] (1.25,-3.5) -- (1.25,0);  
\fill [fill=blue!20,opacity=0.5] (-1.25,0) -- (-1.25,-3.5) arc (180:360:1.25 and 0.5) -- (1.25,0) arc (0:180:1.25 and -0.5);
\draw[thick,blue] (0,0) -- (0,-3.5);
\end{tikzpicture}
\qquad \rightsquigarrow \qquad 
    \begin{tikzpicture}
    [scale=2,baseline={([yshift=0cm]current bounding box.center)}]
    \begin{scope}[x={(1cm,0cm)}, y={(.3cm,.3cm)}, z={(0cm,1cm)}]
        \draw[fill=blue!20,opacity=0.5] (-1.5,-1.5,0) -- (1.5,-1.5,0) -- (1.5,1.5,0) -- (-1.5,1.5,0) -- cycle;
        \draw[->] (-1.7,0,0) -- (1.7,0,0) node[right] {$\tau$};
        \draw[->] (0,-1.7,0) -- (0,1.7,0) node[above right] {$x$};
        \draw[->] (0,0,-0) -- (0,0,1.25) node[above] {$z$};
        \fill[blue] (-1,0,0) circle (1.5pt);
        \fill[blue] (1,0,0) circle (1.5pt);
        \foreach \angle in {0,1,...,180} {
            \coordinate (p\angle) at ({cos(\angle)},0,{sin(\angle)});
        }
        \draw[thick,blue] (p0) \foreach \angle in {10,20,...,180} { -- (p\angle) };
    \end{scope}
\end{tikzpicture}
\end{align*}
   \caption{In the AdS/CFT context, a giant graviton is usually thought of as an object at the center of global AdS. By a conformal transformation, we can view the giant as propagating on a geodesic that intersects the planar boundary at 2 points. The latter picture generalizes more readily  to D$p$-brane (non-conformal) holography.}
    \label{fig:AdS}
\end{figure}

In $\mathcal{N} = 4$ SYM, a giant graviton is an operator of the form \eqref{eq:char} 
where $R$ is a single long column. Equivalently, via the state-operator correspondence, this may be viewed as an energy eigenstate in global AdS, see Figure \ref{fig:AdS}. In preparation for finding the solution in the non-conformal cases, we perform a conformal transformation to Poincare coordinates where the brane propagates between two points on the boundary (along an AdS geodesic which is a semi-circle). We interpret this classical solution as contributing to the 2-pt function of the operator on the plane \eqref{2pt}.%

We propose that the giant gravitons in this context are D$(6-p)$-branes. Recall that the $(p+1)$-form RR field strength is magnetically dual to an $(8-p)$-form field strength, or a $(7-p)$-form potential. The potential is given by\footnote{The case $p=3$ requires special treatment since $C_4$ that is self-dual $C_4\sim \d t \wedge \d\hat\Omega_3+\d\phi \wedge \d\Omega_3$.}  %
\begin{align}
F_{8-p} &= \star F_{2+p} = k_{8-p} \ell_s^{7-p} N \d \Omega_{8-p},\\
C_{7-p} & =  \frac{k_{8-p}}{7-p} \ell_s^{7-p} N \sin^{7-p} \theta \d \phi \wedge \d \Omega_{6-p},\\
k_{8-p} &= (7-p) (2 \pi)^{p-2} d_p.
\end{align}
We will consider an ansatz where the D$(6-p)$ branes homogeneously wrap an $S_{6-p} \subset S_{8-p}$ and move on some trajectory $(z(\tau),\phi(\tau))$, while sitting at some constant $\theta$:
\begin{align}
	\d \Omega_{8-p}^2 = \cos^2 \theta \,  \d \phi^2 +  \sin ^2 \theta\, \d \Omega_{6-p}^2   + \d \theta^2
\end{align}
Pictorially, we are considering a trajectory
\begin{align}
\begin{tikzpicture}[scale=0.95, baseline={([yshift=0cm]current bounding box.center)}]%
\draw[black, thick, rotate=90, fill=gray!20]   plot[smooth,domain=-2:2] (\x, {4-(\x)^2}) ;
\draw[black, thick, fill=gray!20]  (0,0) ellipse (0.5 and 2); 
\fill[blue]  (-0.37,1.34) ellipse (0.1);
\draw (-0.05,1.0) node[blue] {$\tau_2$};
\draw (-1,-0.9) node[blue,scale=0.8] {$\ell$};
\draw[blue, thick, rotate=90]  plot[smooth,domain=0:0.95*sqrt(2)] (\x, {4-2*(\x)^2});  %
\draw[blue, dashed, rotate=90]  plot[smooth,domain=-1.04*sqrt(2):0] (\x, {4-2*(\x)^2});  
\fill[blue] (.33,-1.47) ellipse (0.09) ;
\draw (.2,-1.05) node[blue] {$\tau_1$};
\end{tikzpicture}
\qquad \times \qquad 
\begin{tikzpicture}[scale=1, baseline={([yshift=0cm]current bounding box.center)}]
  \shade[ball color=gray!20, opacity=0.6] (0,0) circle (2);
  \draw[thick] (0,0) circle (2);
  \def\phi{30}
  \coordinate (centerLat) at (0,{2*sin(\phi)});
  \pgfmathsetmacro{\xRadius}{2*cos(\phi)}      %
  \pgfmathsetmacro{\yRadius}{0.35*cos(\phi)}    %
   \fill[blue] (0,{2*sin(\phi)-\yRadius}) circle (2pt);
    \draw[->, blue, thick] (0,{2*sin(\phi)-\yRadius}) -- ++(0.6, 0);
  \draw[blue, thick,dashed, domain=0:180]
       plot ({ \xRadius*cos(\x) },
             { 2*sin(\phi) + \yRadius*sin(\x) });
  \draw[blue, thick, domain=180:360]
       plot ({ \xRadius*cos(\x) },
             { 2*sin(\phi) + \yRadius*sin(\x) });
\end{tikzpicture}
\label{eq:cigar}
\end{align}
The left diagram depicts the Euclidean cigar which describes the black brane solution \eqref{gzz}. The right diagram depicts the 2-dimensional $\phi, \theta$ coordinates where the brane rotates. We are suppressing the $S_{6-p}$ where the brane is wrapped homogenously as well as the spatial directions of the boundary theory (the spatial directions parallel to the $Dp$-brane worldvolume.)
Then the appropriate brane action (in Lorentzian signature) with tension $g_sT_{6-p}=(2\pi)^{-{6+p}} l_s^{-7+p}$ is 
\begin{align} \label{DBI1}
I &= -g_s T_{6-p} \int_{\mathrm{D} ({6-p)} } \d^{7-p} x \, (e^{-\phi} \sqrt{-g_{\mathrm{D}(6-p)}} - C_{7-p} )\\
&=-N  \int \d t \sin^{6-p}\theta \sqrt{  R_\text{AdS} ^2  \left( \frac{h(z)  - h^{-1}(z) \dot{z}^2-\dot{\vec{x}}_p^2}{z^2} \right) -\cos^2\theta \dot\phi^2 - \dot{\theta}^2 } +  N\int \d t \sin^{7-p}\theta \dot \phi.
\label{eq:giant_dbi}
\end{align}
We will consider solutions where $\theta=\mathrm{constant}$, $\phi=\phi(t)$ and $z=z(t)$. Since the Lagrangian is invariant under $\phi \to \phi + a $, the quantity (one of the SO($9-p$) $R$-charges)
\begin{align}
\label{eq:rcharge}
    j=\frac{J}{N}=\frac{ \sin^{6-p} \theta \cos^2 \theta \dot \phi}{ \sqrt{ R_\ads^2 \, \d s^2_\text{AdS}/\d t^2 -\cos^2 \theta \dot\phi^2 }} + \sin^{7-p} \theta
\end{align}
is conserved. If we were to use the proper time along the worldvolume as the time coordinate, we could replace $R_\ads^2 \d s^2_\text{AdS}/\d t^2 \to 1$. This action would then agree with the equation (16) of \cite{Grisaru:2000zn}. %
Here the angular momentum of the brane is related to the operator via \eqref{multi_trace}.
We can then solve for $\dot\phi$ in terms of $j$ and define the generator of proper time translation $\mathcal{H} = J \dot\phi -\mathcal{L}$. %
This also defines an effective action where $\phi(t)$ has been integrated out:
\begin{align}
  -  I_\text{eff}=  M_\text{eff}(j,\theta) R_\ads \int \d s_{\text{AdS}} =  M_\text{eff}(j,\theta) R_\ads \int \d t \sqrt{ \frac{h(z)  - h^{-1}(z) \dot{z}^2}{z^2}  },
\end{align}
This is the action for a point particle in the AdS black brane metric.
 Following \cite{Grisaru:2000zn}, we find that the effective mass is given by
\begin{align}
\label{eq:Meff}
    M_\text{eff}(\theta,J) &= N \sqrt{j^2 + \tan^2 \theta (j- \sin^{5-p} \theta)^2}.%
\end{align}
We plot the effective mass above the BPS bound ($M_{\mathrm{eff}}=J$) below for the D0 brane background:
\begin{align}
\includegraphics[width=.7\columnwidth,valign=c]{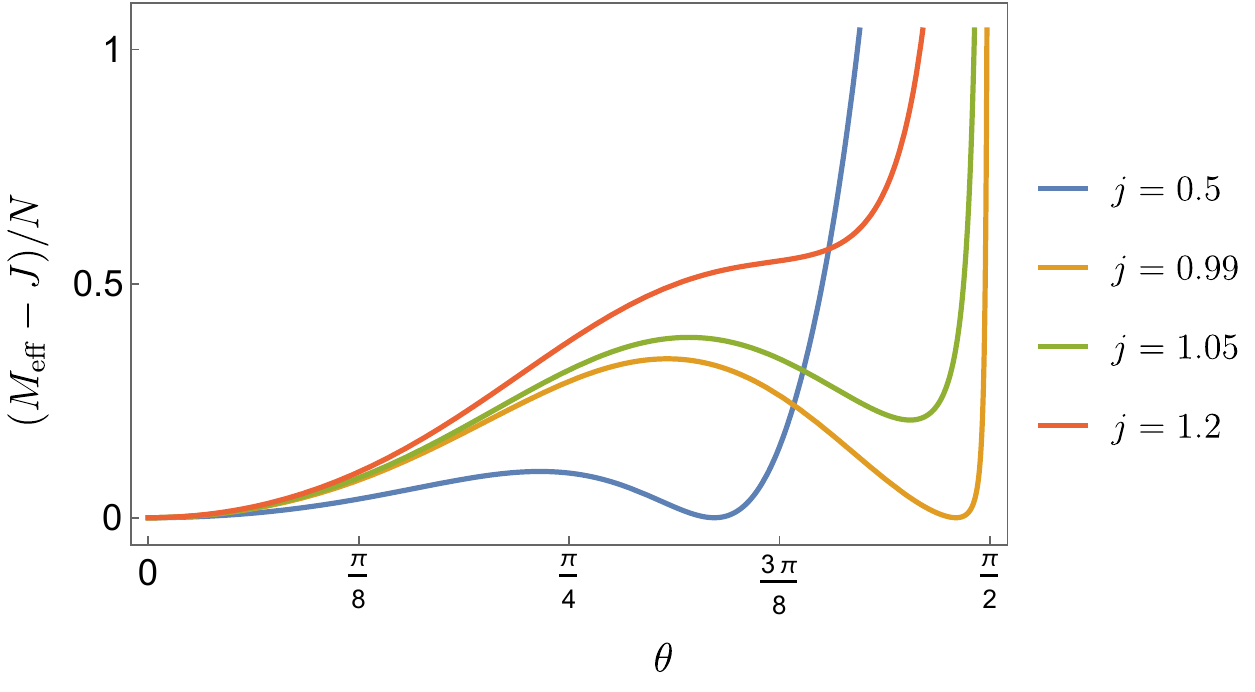}
\end{align}
As $j$ approaches 1, the effective potential develops a corner at $\theta = \pi/2$. The giant is becoming maximal. 
In the right figure of \eqref{eq:cigar}, the maximal giant sits at the north pole but wraps a maximal $S_{6-p}$. 
For fixed angular momentum $j$, minimizing $M_\text{eff}$ with respect to $\theta$ gives\footnote{One can also find anti-brane solutions. These are described by $-A_{7-p} \to + A_{7-p}$ in \eqref{DBI1}. One obtains $M_{\mathrm{eff}}(\theta,J)=N\sqrt{j^2+\tan^2\theta(j+\sin^{5-p}\theta)^2}$; see \cite{Grisaru:2000zn}.} $\sin \theta_\text{min} = j^{1/(5-p)}$ for $j<1$. We plot the minimum $M_\text{eff}(\theta_\text{min}, J)$ here for the $p=0$ and $p=1$ case:
\begin{align}
\includegraphics[width=.7\columnwidth,valign=c]{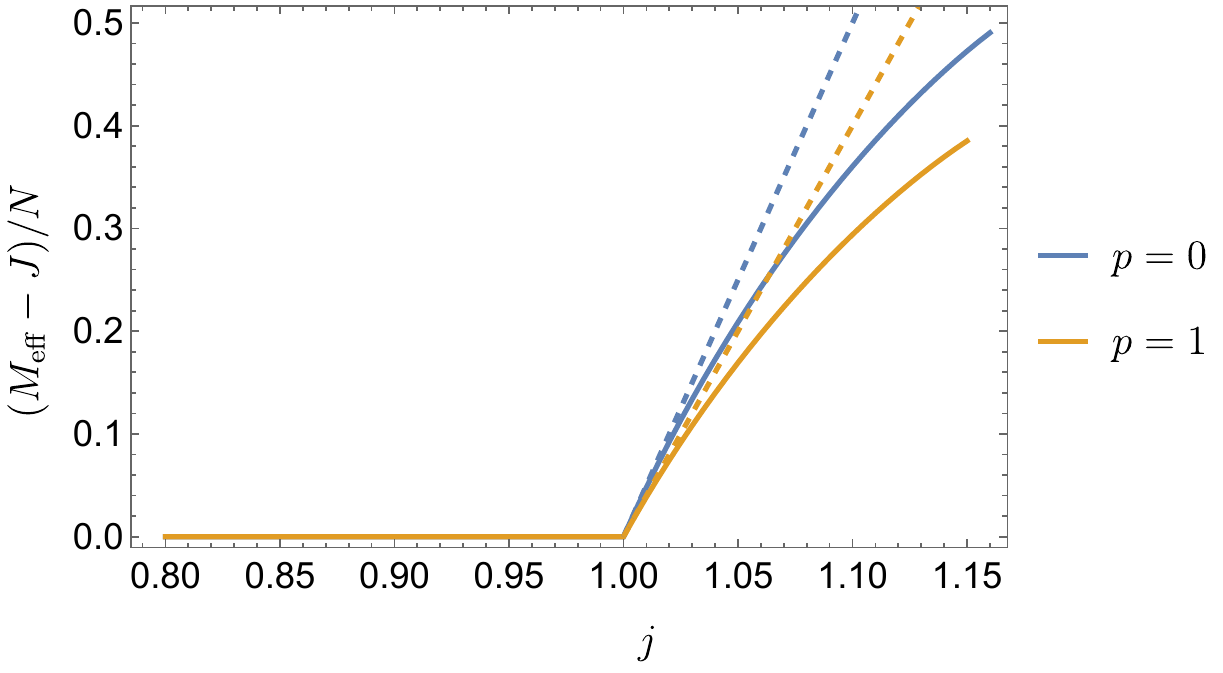} \label{bps-nonBPS}
\end{align}
We see that for $j\leq 1$ and $p<5$, the giant saturates the BPS bound, satisfying $M_{\mathrm{eff}}=J$.
For $j>1$, the non-trivial minimum no longer saturates the BPS bound \cite{Grisaru:2000zn}, and for large enough $j$ the second minimum is lost. This is the stringy exclusion principle \cite{Maldacena:1998bw}.
Note the non-trivial BPS solutions for $j>1$ do {\it not} correspond to the operators $\chi_R(\mathcal{Z})$. For $j = 1^+$ these solutions have $\dot\phi = 6-p$ as opposed to $\dot \phi = 1$ for $j<1$. These do not violate the speed of light bound for the expanded brane configurations, $\cos^2\theta  \dot\phi^2<1$. 
In Appendix \ref{app:quantum}, we show that the slope of the dashed lines displayed in \eqref{bps-nonBPS} are given by $2/R_\ads = 5-p$, see \eqref{eq:energy_above_BPS}.

For the giant graviton solutions, $M_\text{eff}=J$ is the classical extrapolation of the scaling dimensions of the operators \eqref{eq:dim}:
\begin{align}\label{eq:dimj}
	\Delta = M_\text{eff} R_\text{AdS} = \frac{2}{5-p} J .
\end{align}
We view this agreement as strong evidence for our proposal that the giants are D$(6-p)$ branes\footnote{Note that our formulas are singular at $p=5$; the above analysis is therefore limited to $p<5$. We discuss the $p=5$ and $p=6$ cases in Appendix \ref{app:higher_p}; the giant gravitons seem to be quite different in these cases.}. %
Note that the trajectory of the giant in the radial direction is not a geodesic in the original metric \eqref{gzz}. That geodesic {\it would} be the trajectory followed by a massive string state; but the giant instead follows a trajectory given by the ``fake'' AdS. This is expected, since the geodesic with respect to the original metric \eqref{gzz} leads to a non-conformal 2-pt function.

Our classical analysis shows that the giant gravitons have  dimensions \eqref{eq:dimj}, but one could ask if the near-maximal giants have precisely the correct BPS spectrum; e.g., $J$ quantized to take integer values $\le N$ and precisely one state for each value of $J$, since this is required from the boundary analysis of $\hf$-BPS operators for any value of $p$. In Appendix \ref{app:quantum}, closely following \cite{Lee:2024hef}, we analyze part of the spectrum of fluctuations of the giants (expanded around the maximal giant) {and show the quantization of $J$}, modulo an assumption about other modes of the D($6-p$)-brane. This is enough to reproduce the second term in \eqref{eq:sum} for general $p$.

We have emphasized the scaling dimensions \eqref{eq:dimj} which imply that the 2-pt function of giant graviton operators at low temperatures has the conformally invariant form \eqref{2pt}. At finite temperature (while staying in the strongly coupled 't Hooft regime), the connection to the AdS black brane also allows one to obtain the thermal 2-pt function of giant graviton operators. This is particularly interesting because the thermal 2-pt function probes the black hole geometry; we explore this in Appendix \ref{app:thermal}.

\section{Extrapolation to strong coupling \label{sec:discuss}}

In this section we discuss what happens when the giant gravitons enter the strong coupling region for $-1\le  p \le 4$. In the IIA cases we should uplift to the M-theory description whereas in the IIB cases we should use the S-dual description. 
For the $p>4$ case, the UV completion of the SYM theory is rather exotic and the decoupling limit is subtle \cite{Itzhaki:1998dd}. We relegate a discussion of these cases to the appendices.

\subsection{D(-1)-branes/IKKT \label{sec:ikkt}}

The case of $p=-1$ D-instantons/IKKT requires special attention as there is no space or time on the boundary. The reader may wish to read subsections (\ref{BFSS}-\ref{6D}) first before returning to this subsection.

Unlike the higher dimensional cases, there is no Euclidean time, and hence no temperature parameter. Therefore the dual geometry is always related to Poincar\'e-AdS.
Although there is no boundary spacetime, there is a family of classical giant graviton solutions, which correspond to a D7 brane that starts at the boundary $z=0$ and falls radially inward in the $z$ direction, while rotating on the $S_9$, e.g., $\phi(z)$ determined by the angular momentum $J$.

Before discussing the interpretation of this classical solution, let us comment on another issue. Eventually the D7 brane enters the region of strong coupling. To address this, first recall that Type IIB string theory enjoys an $SL(2,\mathbb{Z})$ symmetry:
\begin{align}
    \tau \to \frac{a \tau + b}{c \tau + d}, \quad \bar\tau \to \frac{a \bar{\tau} + b}{c \bar{\tau} +d },
\end{align}
where the integers $a,b,c,d$ satisfy $ad-bc = 1$ and $\tau= A + e^{-\phi}$ is the Euclidean modular parameter. With the convention that the axion $A$ is real, $\tau$ vanishes on this background whereas %
\begin{align}
    \bar\tau &= A - e^{-\phi} = -2 \frac{d_{-1}  N}{(2\pi)^{3}} \left( \frac{z}{R_\ads} \right)^{-8/3}.%
\end{align}
Under an $SL(2,\mathbb{Z})$ transformation, the D7 brane that we are discussing transforms into an exotic 7-brane sometimes referred to as  a $(p,q)$ 7-brane \cite{Bergshoeff:2006gs}. The action for this brane is an $SL(2,\mathbb{Z})$ invariant given by \cite{Eyras:1999at, Bergshoeff_2005, Bergshoeff:2006gs}  %
\begin{align}
    \mathcal{L}&=q_\alpha q_\beta\mathcal{M}^{\alpha\beta}\sqrt{\det(g_E)}+q_\alpha q_\beta \mathcal{C}^{\alpha \beta}_{8}, \quad  \mathcal{M}_{\alpha \beta} =e^\phi\begin{pmatrix}
        A^2+e^{-2\phi} &  A\\
        A & 1
    \end{pmatrix}
\end{align}
Here $g_E$ is the Einstein frame metric and $q_\alpha$ is an 
$SL(2,\mathbb{Z})$ doublet that characterizes the brane, e.g. $q_\alpha=(0\  -1)^T$ corresponds to a D7 brane. The second term in the action reduces to $-\int C_8$ for a D7 brane, for
$\d C_8^{\alpha\beta}=F^{\alpha\beta}_9$, with $F^{\alpha\beta}_9$ being the Hodge dual of the 1-form in Eq. (2.3) of \cite{Bergshoeff_2005}.

Because the action of the brane is an $SL(2,\mathbb{Z})$ invariant, we can say that the $(p,q)$ brane still ``feels" like it is in AdS in any $SL(2,\mathbb{Z})$ frame. Ideally one would like to find a frame in which the $z \to \infty$ region is weakly coupled but we did not find such a frame\footnote{Consider the transformation given by $a=k+1,\ b=1,\ c=k,\ d=1$, for large $k$. This takes $e^\phi \to e^\phi - 2k$, which is $\mathcal{O}(\epsilon)$ in a small region around $z=z_0$, for $z_0=(2k)^{3/8}$. However, the size of this region is $\mathcal{O}(\epsilon z_0^{-5/3})$ and the coupling grows very rapidly, so this transformation is not reliable. Note the naive transformation $(\tau \to -1/\tau,\bar \tau \to -1/\bar \tau)$ does not work since $\tau=0$.}.

Let us now comment on the interpretation of the giant graviton solution in the IKKT model \cite{Ishibashi:1996xs}.
Naively, one might try to interpret this classical solution as a contribution to the 1-pt function of $\chi_R(\mathcal{Z})$, but an immediate objection is that such a 1-pt function automatically vanishes since the IKKT model is SO(10) invariant. One can also consider a 2-pt function of $\ev{\chi_R( \mathcal{Z})  \chi_R(\bar{\mathcal{Z}}})$, but based on the analogy with AdS\footnote{The analog in higher dimensions would be computing a 2-pt function with vanishing separation in spacetime. The analog of the radially inward trajectory here would be a 2-pt function, with one of the points at spatial infinity on the boundary.}, this 2-pt function would be dominated by configurations where the brane is in the UV region. 
Another issue with this solution is that we did not find an $SL(2,\mathbb{Z})$ frame where the string coupling is small deep in the radial direction. However, we believe that all of these issues could be remedied in the more controlled context of the mass-deformed IKKT model \cite{Hartnoll:2024csr, Komatsu:2024bop, Komatsu:2024ydh}. In that model, the SO(10) symmetry is broken so the giant graviton operators could have non-trivial 1-pt functions. %
Note that the mass deformation is relevant and modifies the strongly coupled region while keeping the asymptotic region intact \cite{Komatsu:2024bop}. One might speculate that the giant graviton solution that we found would still approximately describe the trajectory of the D7 brane which is ``close'' to the boundary, but deep in the radial direction the solution would be modified. We hope to explore these issues more concretely in the future.

\subsection{D0-branes/BFSS \label{BFSS}} 
The 11d uplift of the D6 brane is a Kaluza Klein monopole. If we consider a low temperature limit \cite{Itzhaki:1998dd} (beyond the standard 't Hooft regime) $N^{5/9} \gg \lambda^{1/3} \beta \gg N^{10/21}$, and we consider separated points  $\tau \propto \beta$ on the Euclidean circle, we can arrange for the D6 branes to pass into the M-theory region. In the vicinity of the Kaluza Klein monopole this should be described by a metric of the form
\begin{align}
	\d s^2_{11} \sim - \d t^2 + \d x_i^2 + \d s^2_\text{Taub-NUT}.
\end{align}
Here $(t,x_i)$ are the coordinates that were parallel to the D6 brane and $\d s^2_\text{Taub-NUT}$ describes a 4-manifold. These 4 directions are associated with the radial direction, $\phi, \theta$ (the two directions on the $S_8$) and the M-theory circle.   
It seems natural to conjecture that the power law \eqref{2pt} holds in this regime since the action of the Kaluza-Klein monopole should reduce to the D6 brane action, see \cite{Bergshoeff:1997gy,Imamura:1997ss}. Note that the black string in M-theory also inherits the scaling symmetry of the Type IIA solution  \cite{biggs2023scaling}.
\subsection{D1-brane/F1-brane \label{d1f1}}
In the D1-brane theory, the giants are D5 branes wrapping an $S_5 \subset S_7$. When we go to the IR of the D1-brane theory, the dilaton grows and $S$-duality relates the D1 black brane to the F1 black brane (the black string) solution \cite{Itzhaki:1998dd}. Under S-duality the giant D5 branes should become NS5 branes. Let us work this out more explicitly. Under S-duality, the Einstein frame metric is invariant, the dilaton $\phi \to -\phi$, and $A \leftrightarrow B$. This gives the F1 black brane solution:
\begin{align}
    \frac{\d s^2}{\alpha'} &=  \frac{d_1 N}{2\pi}\left(   \frac{R_\ads}{z} \right)  \left[R_\ads^2 \left(\frac{h(z)\, \d \tau^2+ h^{-1}(z) \d z^2 +\d x^2}{z^2}\right)+\d \Omega_{7}^2\right],\\
    e^{2\phi} &=   \left( \frac{d_1 N}{2\pi} \right)^{2} \left( \frac{z}{R_\ads} \right)^{-3},\\
    B_{01} &= \frac{\alpha' d_1 N}{2\pi} \left( \frac{z}{R_\ads} \right)^{-3},
\end{align}
where $B$ is the NS-NS 2-form potential. The magnetic dual is
\begin{align}
	\d B_6 &= e^{-2\phi } (\star H_3),  \\ 
    B_6 &=  \frac{d_1 N}{2\pi} {\alpha'}^3 \sin^6\theta  \d \phi \wedge \d \Omega_5.
\end{align}
The worldvolume action is (for $T_{\mathrm{NS5}} = ((2\pi)^5 \ell_s^6)^{-1}$)
\begin{align}
    -I &=  T_{\mathrm{NS5}} \int \left( \d^{6} x\,   e^{-2\phi } \sqrt{-g_\text{NS5}} - B_6\right) \\
    &= N  \int \d t  \left( \sin^5 \theta \sqrt{R_\ads^2\left(\frac{h(z)-h^{-1}(z)\dot z^2-\dot x^2}{z^2}\right)-\cos^2\theta \dot\phi^2} -  \sin^6 \theta \dot \phi \right).
\end{align}
We see that this is the same as the action (\ref{eq:giant_dbi}) of the giant in the D1 brane geometry, and similarly the NS5 brane also ``feels" as if it lives in AdS.
This implies that the 2-pt function \eqref{2pt} continues to hold, even when we consider separations in $x$ that are so large that we have left the 't Hooft regime. This is a non-trivial prediction about the strong coupling regime of the boundary theory.

In the F1 brane background, the dilaton shrinks towards the IR and eventually we are left with a weakly coupled theory in the deep IR. This is the free symmetric orbifold CFT. The leading irrelevant operator is a twist operator \cite{Dijkgraaf:1997vv}. It would be interesting to understand the description of these NS5 branes in this symmetric orbifold theory\footnote{We thank Shota Komatsu and Steve Shenker for discussions on this point.}.

\subsection{D2-branes/ABJM}
In the D2-brane theory, at strong coupling we should use the 11-d uplift which is \cite{Itzhaki:1998dd} $\ads_4 \times \mathrm{S}_7$. On the boundary side, the theory flows to a strongly coupled 3D conformal field theory, the ABJM theory \cite{Aharony:2008ug}.
Then we expect the D4 giant to become an M5 brane, which wraps an $S_5 \subset S_7$ in the internal space of $\ads_4 \times \mathrm{S}_7$.
This makes contact with the M5 brane giant graviton analysis in $\ads_4 \times \mathrm{S}_7$ in \cite{McGreevy:2000cw, Grisaru:2000zn, Hashimoto:2000zp}, see also discussions about giant gravitons in ABJM \cite{Sheikh_Jabbari_2009}. In the IR, the giants feel AdS because they actually live in AdS! 

Note, however, that the ratio $R_{\ads_4}/R_{S_7} = 1/2$ so $\Delta_\text{ABJM} = J/2$ which is different than our formula $\Delta_{\text{D}2} = 2 J/3$ derived in the 't Hooft limit of the D2 brane theory. Thus the 2-pt function of giant gravitons has a different power law in the deep IR than in the ``UV'' ('t Hooft regime) where it is governed by the $p=2$ black brane geometry. 

\subsection{D3 branes}
This is the most familiar case, the giants in $\mathcal{N}=4$ SYM are D3 branes \cite{McGreevy:2000cw} and are self-dual.

\subsection{D4-branes/6d (2,0) theory \label{6D}}
For $p>3$, the Yang-Mills interaction is irrelevant; we view the $p=4$ Yang-Mills theory as an effective field theory with the UV completion given by the 6D (2,0) conformal field theory on a spatial circle (with supersymmetric boundary conditions). In the dual gravity description, we have a IIA regime which should be uplifted to an M-theory solution in the UV. This M-theory solution is $\ads_7 \times \mathrm{S}_4$ and the D2 giants become M2 branes that wrap an $S_2 \subset S_4$ found in \cite{McGreevy:2000cw}. Note that $R_{\ads_7}/R_{S_4} = 2$, which also agrees with $R_\ads = 2/(5-p)$ for $p=4$. Hence the UV dimension of the operator agrees with the one derived in the 't Hooft limit of the D4 brane theory $\Delta = 2 J$. Note that in the UV, we are considering operators in the 6D (2,0) theory that are localized on the circle.

\section{Discussion} %

We list some open questions:
\begin{itemize}
    \item Calculating the spectrum of fluctuations (including non-BPS modes) of these giants and their quantum corrections is a concrete future direction.  In Appendix \ref{app:quantum}, we analyze a simple subset of these fluctuations. Our analysis\footnote{Even at the classical level, one could consider the non-BPS giants with $j>1$ in \eqref{bps-nonBPS} and ask for the boundary duals of such objects.} also implies that there should be many operators in the boundary theory which are dual to non-BPS giants; these operators would exhibit scaling symmetry \eqref{2pt} despite being non-supersymmetric. 
    \item Finding the dual giants for $p \ne 3$ is an interesting challenge. For $p=3$ the dual giants wrap the AdS sphere. This picture is most clear in global AdS. Without a known analog of global AdS in the non-conformal cases, it seems tricky to find the dual giants for $p\ne 3$. Nevertheless, there are certainly operators $\chi_R(\mathcal{Z})$ where $R$ is a long row (a fully symmetric irrep) so presumably acting with these operators must create {\it something} in the bulk and it seems likely that the dual giants should also ``feel'' AdS based on the extrapolation from small gravitons. It would also be nice to find the analog of the LLM geometries \cite{Lin:2004nb} which  describe $\chi_R(\mathcal{Z})$, where $R$ is a more general Young diagram.

    \item One could study the giants in the BMN matrix model \cite{Berenstein:2002jq}. The BMN matrix model has vacuum states that contain M2 and M5 branes \cite{Berenstein:2002jq,  Dasgupta:2002hx, Dasgupta:2002ru, Maldacena:2002rb}. Our analysis suggests that there should also be states in the BMN theory that contain KK monopoles by acting with operators like $\chi_R(\mathcal{Z})$. Perhaps localization \cite{Asano:2014vba, Asano:2017xiy} might be applicable for computing such correlators of such operators, see also \cite{Bobev:2019bvq, Bobev:2024gqg}.
    \item As discussed further in Appendix \ref{app:higher_p}, we raise some unresolved puzzles about giant gravitons in $p=5$ and $p=6$. We did not find a clear candidate for the bulk dual of the "tall column" operators in these cases. 
    \item As mentioned in \ref{sec:ikkt}, the interpretation of the classical giant graviton solutions for the D-instanton/IKKT context requires clarification. It seems natural to wonder about giant gravitons in the mass-deformed IKKT model \cite{Hartnoll:2024csr, Komatsu:2024bop, Komatsu:2024ydh} where perhaps localization might also give useful information from the boundary. On a separate note, as with the D1 black brane solution, the dilaton in the S-dual picture \eqref{eq:s_dilaton} shrinks as we go deep in the radial direction; it would be interesting if there is a free theory that describes the deep infrared of the IKKT model, analogous to the free orbifold CFT in the D1-brane case. The meaning of ``infrared'' in IKKT is unclear but in the case of mass-deformed theory there may be such a notion.
    \item $n$-pt functions of giant graviton operators: it would be interesting to compute higher point correlation functions of giants/super-graviton operators and ask whether they are conformally covariant (and not just scale invariant). %
    We could ask whether the scaling symmetry of these supergraviton modes/giants is enhanced to a supersymmetric algebra, which would explain the relation $M \ge J$. %
 {$n$-point functions of giant graviton operators have been considered in the AdS context from both the bulk and boundary points of view in for instance \cite{Corley:2001zk,Hirano:2012vz,Holguin_2023}.}
 
\end{itemize}

\section*{Acknowledgments}
We thank Yiming Chen, Alexander Frenkel, Shota Komatsu, Ji Hoon Lee, Juan Maldacena, Jo\~ao Penedones,  Stephen Shenker and Douglas Stanford for useful discussions. HL is supported by a Bloch Fellowship and by NSF Grant PHY-2310429.

\appendix

\section{Spectrum of the giant graviton \label{app:quantum}}
In this section we give a partial analysis of the perturbative spectrum of the giant graviton (around the maximal giant), following closely the work of \cite{Lee:2024hef}, see also \cite{McGreevy:2000cw, Gautason:2024nru, Beccaria:2024vfx}.

\subsection{Classical analysis}
Let's reconsider the action for the giant graviton:
\begin{align}
   I &= N \int \d \tau \left[ -   \sin^{6-p}(\theta) \sqrt{  1 -\cos^2\theta \dot\phi^2 - \dot{\theta}^2 } +    \sin^{7-p}\theta \dot \phi\right].
\end{align}
To work out the full spectrum of fluctuations we should also include other scalar, vector and fermionic modes. However at least in the D3 case \cite{Gautason:2024nru}, exciting these other modes lead to non-BPS states so we will not consider them here.
Here $\dot{f} = \d f/\d \tau$ is a time derivative with respect to written proper time measured with respect to the fictitious AdS metric,
\begin{align}
   \left(\frac{\d \tau}{\d t}\right)^2 =   R_\text{AdS} ^2  \left[ \frac{h(z)  - h^{-1}(z) z'(t)^2-{\vec{x}_p'(t)}^2}{z^2} \right].
\end{align}
Now, following \cite{eleftheriou2025localizationwallcrossinggiantgraviton,Lee:2024hef, Lee:2023iil, McGreevy:2000cw} we zoom in on the maximal giant and write
\begin{align}
    \theta = \frac{\pi}{2} - \frac{r}{\sqrt{N}}.
\end{align}
Defining $\Pi = \dot{r}$, to quadratic order the $R$-charge is
\begin{align}
L &= N(-1+\dot{\phi})  + \frac{\dot{r}^2 + r^2 (\dot\phi)^2 }{2} + \frac{6-p}{2} r^2 - \hf (p-7) r^2 \dot \phi,\\
J &= N- \hf(p-7) r^2 + r^2 \dot \phi, \\
H &= \dot \phi J + \dot{r} \Pi - L  \\
&=  N +\hf (N-J) (p-7)  +\frac{ \Pi^2 +  (\frac{5-p}{2})^2 r^2 }{2}    + \frac{\left(N- J\right)^2}{2 r^2}\\
&= J + \frac{J-N}{R_\ads}+ \frac{ \Pi^2 +   (r/R_\ads)^2 }{2}    + \frac{\left(N- J\right)^2}{2 r^2}.%
\end{align}
Note that the Hamiltonian is conjugate to the proper time in AdS; it is essentially the mass of the giant. Now after a canonical transformation, we see that %
\begin{align}
    \tilde{r} &= r \sqrt{R_\ads}, \quad \tilde{\Pi} = \Pi/\sqrt{R_\ads},\\
     H-J &=   \frac{1}{R_\ads}\left[  \frac{ \tilde\Pi^2 }{2}    + \frac{\left(N- J + r^2\right)^2}{2\tilde{r}^2} \right].
\end{align}
The RHS is the sum of two positive terms, so $H-J \ge 0$. Furthermore, the BPS condition $H-J = 0$ imposes $\tilde \Pi = 0$ and $J = N - \tilde{r}^2$.

\subsection{Quantum analysis}
Just as in \cite{Lee:2024hef}, we can map to a 2D harmonic oscillator in polar coordinates;
\begin{align}
\ell &= J-N,\\
H-J &= \frac{1}{R_\ads} \left( h+\ell \right),\\
h &= \frac{\tilde\Pi^2 + \tilde{r}^2}{2} + \frac{\ell^2}{2 r^2}.
\end{align}
This allows us to quantize the system easily. We get essentially the same results as  \cite{Lee:2024hef} except that $H-J$ has a spectrum that is evenly spaced in units of $1/R_\ads$. More explicitly, let $n_1, n_2 \in \mathbb{Z}_{\ge 0}$, then the spectrum is %
\begin{align}
    H-J = \frac{1}{R_\ads} (2n_2+E_0) \label{eq:energy_above_BPS},\\
    J = N + n_2 - n_1.
\end{align}
Here $E_0 = 1$ is the vacuum energy of a single harmonic oscillator. As in \cite{Lee:2024hef}, the value of $E_0$ is shifted due to the presence of other fermionic and bosonic and fermionic modes on the brane. In the $\mathcal{N}=4$ SYM case, the results of \cite{Beccaria:2024vfx} showed that accounting for these modes effectively shifts $E_0 \to 0$; we conjecture that a similar analysis for the D$(6-p)$ branes would also give $E_0 \to 0$.
The $n_2= 0$ solutions correspond to the BPS giants; the ones with $n_2 > 0$ correspond to the non-BPS giants, see \eqref{bps-nonBPS}.

The arguments in \cite{Lee:2024hef} can then be repeated; in particular one can use their phase space path integral to see that there is a contribution to the partition function %
\begin{align}
    \Tr_\text{BPS} q^R  \supset -\frac{q^{N+1}}{1-q},
\end{align}
where the important minus sign is interpreted as coming from the trace relations; from the path integral point of view the maximal giant is an unstable saddle point (we refer readers to the discussion around (3.17) in \cite{Lee:2024hef}.)
The main point we are making here is that at this level of approximation, the bulk problem is essentially uniform in spacetime dimensions $p$.

Let us also comment on the interpretation of the BPS partition function. In  $\mathcal{N} = 4$ SYM we can view this partition function as a trace over (a BPS sector of) the Hilbert space of $\mathcal{N}= 4$ SYM on $S^3$. Without the state-operator correspondence, we do not have a similar interpretation for $p\ne 3$. Nevertheless, we can interpret this as a trace over the bulk particles in the theory, or equivalently as counting BPS operators in the boundary theory.

\section{Giant gravitons for \texorpdfstring{$p=5$}{p=5} and \texorpdfstring{$p=6$}{p=6} \label{app:higher_p}}
\subsection{\texorpdfstring{$p=5$}{p=5}}
For $p=5$ we cannot use the $z$ coordinates since the coordinate transformation to them is singular. We work instead with the metric
\begin{align}
    \d s^2&= \rho(-h(\rho)\d\tau^2+\d x_5^2)+h^{-1}(\rho)\rho^{-1} \d\rho^2+\rho \d\Omega_{3}^2,\\
    h(\rho)&=1-\frac{\rho_0^2}{\rho^2},\\
    e^{-2\phi} &= (d_5 (2\pi)^3 N)^2 \rho^{-2},\\
    A_{0 \cdots 5}&=(\sqrt{\alpha'})^{6}d_5(2\pi)^3 N \rho^{2}.
\end{align}
The action for the giant graviton becomes
\begin{align}
    I &= -\frac{1}{2\pi l_s^2} \int_{\mathrm{D} {1} } (\d^{2} x \, e^{-\phi} \sqrt{-g_{\mathrm{D}1}} - A_{2} )\\
    &= -N \int \d t \sin \theta \sqrt{h-{\dot{y}^2}/h-\dot{x}^2-\cos^2\theta \dot\phi^2 } +N \int \d t \sin^2\theta \dot \phi,\\
    y &= \log \rho/\rho_0. 
\end{align}
Here we have defined a new coordinate $y$; in the extremal limit $h=1$, we see that the D1 brane feels like it lives in Minkowski space.
Note that this action still respects the scaling similarity in \cite{biggs2023scaling}, where it is interpreted as a rescaling of the radial direction $\rho$. We can follow the same procedure as in Section \ref{sec:solution} and compute the effective action where $\phi(t)$ is integrated out:
\begin{align}
    I_\mathrm{eff}=M_\mathrm{eff}(j,\theta) \int dt \sqrt{h-\frac{\dot{y}^2}{h} -\dot{x}^2},
\end{align}
where $M_\mathrm{eff}(j,\theta)$ is given by Eq. (\ref{eq:Meff}).
 However, there are no stable giant graviton solutions since $M_\mathrm{eff}(j,\theta)$ does not have a minimum except at $\theta=0$ (corresponding to the pointlike graviton). This seems reminiscent of a puzzle in AdS$_3$ $\times$ S$_3$ $\times$ M$_4$, where there is also no non-trivial minimum for the D$1$ brane, as discussed in \cite{McGreevy:2000cw}. In future work, it would be interesting to understand what the bulk description is of the tall column operators in this case.

In the $p=5$ case we can view SYM as the low energy effective field theory of ``little string theory'' \cite{Itzhaki:1998dd, Aharony:1998ub}. The strong coupling region is described by the NS5 black brane solution. The candidate giants would then be F1 brane (strings).

\subsection{\texorpdfstring{$p=6$}{p=6}}
For $p=6$, the giants are D$0$ branes, and the DBI action evaluates to
\begin{align} \label{lagrange0}
I =- \frac{N}{2}  \int \d \tau   \sqrt{  1 -\cos^2\theta \dot\phi^2 - \dot{\theta}^2 } -  \sin\theta \dot \phi.
\end{align}
Note there is an extra $1/2$ multiplying the whole action. This is the Lagrangian of a charged relativistic particle on a 2-sphere in the presence of a magnetic monopole\footnote{We thank Igor Klebanov for discussions related to this point.}. 
This setup is SO(3) symmetric and we can write the Hamiltonian
\begin{align}
    H &= \sqrt{m^2 + \vec{J}^2 - q^2} =  |\vec{J}|,\\
    \vec{J}^2 &= \mathsf{j}(\mathsf{j}+1) , \quad \mathsf{j} = q, q+1, q+2, \cdots ,\\
    m &= q = N/2.
\end{align}
In the second line we have quantized the theory, along the lines of \cite{Haldane:1983xm}.
We see that all states in the quantum theory are BPS states. Ignoring the fermions, we find $2 \mathsf{j} + 1$ states for each value of $\mathsf{j}$. In the classical approximation, the minimum value of $\mathsf{j} = q$ corresponds to a D0 brane sitting at rest (not rotating) on the $S_2$ at a location $\vec{r} \propto \vec{J}$; the larger values of $\mathsf{j}$ correspond to a D0 moving on the sphere in a circular trajectory (the higher Landau levels). For $\mathsf{j}$ near the minimum value of $q$, the problem is essentially non-relativistic and we recover the results of \cite{Haldane:1983xm}, but the relativistic corrections are important in showing that all states are BPS.

 This implies the existence of BPS states with $j>1/2$, or a pair of D0-anti-D0 branes which have $j>1$. It is not clear how to interpret these solutions in terms of boundary operators; we leave this puzzle for future work. The D0 $+$ anti-D0 brane on an $S_2$ is very similar to the analogy for giants proposed in \cite{McGreevy:2000cw}. However, a key difference in this setup is that the D0 branes have a mass/charge ratio of 1, whereas \cite{McGreevy:2000cw} considered charged particles with mass/charge ratio $\approx 0$.

Note that for $p=6$, the D6 worldvolume theory does not decouple from the bulk \cite{Itzhaki:1998dd}.  
According to the bulk solution, the UV region is described by M-theory \cite{Itzhaki:1998dd} where the D0 brane would uplift to an 11d graviton. We leave a more thorough analysis of this case for the future.

\section{Thermal 2-pt function}
\label{app:thermal}

\begin{figure}
    \centering
     \includegraphics[width=0.7\columnwidth]{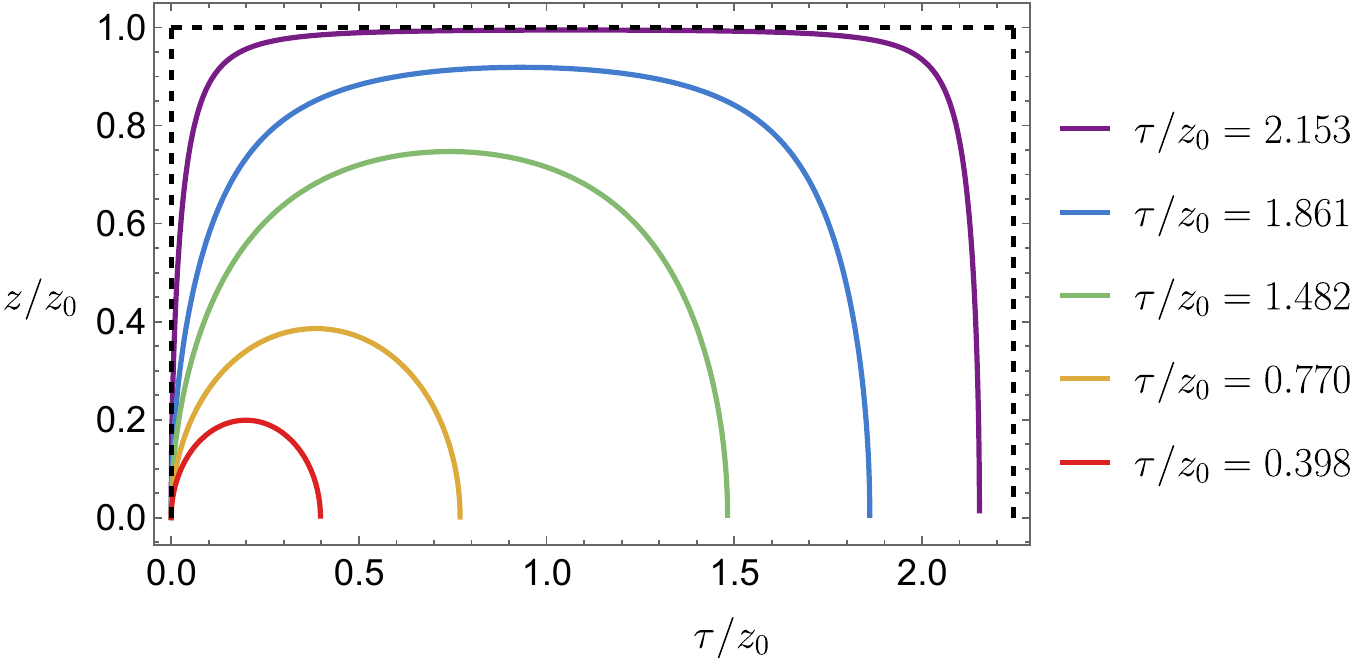}
     \caption{Trajectories of the giant graviton in the $z-\tau$ plane, for the finite temperature black hole $p=0$. For small separation, we see that the trajectories are approximately semi-circles, which is what we expect from vacuum AdS. So for $\tau \ll \beta$ we get power law behavior in the correlators. Here $\beta/z_0 =\frac{10\pi}{7}$; as $\tau \to \beta/2$ from below, the $z$ trajectory approaches the dotted black lines. The actual trajectory is smooth on the Euclidean cigar; see \eqref{eq:cigar} where the curve passes through the horizon (the tip of the cigar). 
     }
     \label{fig:geodesic}
 \end{figure}

At finite temperature, the relation to AdS$_{d+1}$ black brane also gives a simple prescription to find the large $N$ thermal 2-pt function $\ev{\bar{O}_\text{BPS}(\tau_{1}) O_\text{BPS}(\tau_{2})} \sim e^{-I}$ by evaluating $I$ on a geodesic in the Euclidean AdS black brane%
\begin{align}
    \ev{\bar{O}_\text{BPS}(\tau_{1}) O_\text{BPS}(\tau_{2})} \approx \exp \left[ {-\Delta \int \d \tau \sqrt{\frac{h(z) +h^{-1}(z)  {\dot z}^2}{z^2}}} \right].
\end{align}
To solve the geodesic equation, we introduce the conserved quantity
\begin{align}
    E=-\frac{h(z)}{z \sqrt{h(z) + h^{-1}(z) \, \dot{z}^2}}.
\end{align}
We can then compute the renormalized length
\begin{align}
    \tilde{\mathcal{L}}=\lim_{\epsilon\to 0} \left[2\int_\epsilon^{z_*} \frac{\d z}{z \sqrt{h(z)-E^2z^2}}+2\log \epsilon\right],
\end{align}
where $z_*$ is the turning point of the geodesic in the bulk where it satisfies $\dot z=0$. %
We plot the geodesics in Figure \ref{fig:geodesic}. Due to the scaling symmetry of the planar AdS black brane, the trajectory $\tilde{z}(\tilde{\tau})$ is actually independent of $\beta$ when written in terms of $\tilde{z} = z/z_0$ and $\tilde{\tau} = \tau/z_0$.%

When the two points are located on diametrically opposite ends of the thermal circle, we can analytically find the length of the spacelike geodesic joining them, $\tilde{\mathcal{L}}=\frac{4}{d}\log2+2\log z_0$. This indicates that the associated two-sided 2-pt function behaves like a power law, $e^{-\Delta \tilde{\mathcal{L}}}\sim z_0^{-2\Delta}$, again signaling the scaling symmetry.

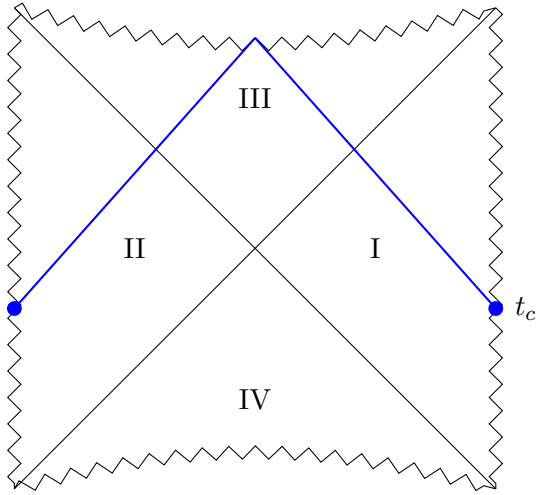
\begin{figure}[H]
\centering
\begin{tikzpicture}[scale=0.8]
\node (I)    at ( 4,0) {};  %
\node at (2,0) {I};
\node at (-2,0) {II};
\node (II)   at (-4,0) {}; %
\node (III)  at (0, 2.5) {III};
\node (IV)   at (0,-2.5) {IV};
\path  
  (II) +(90:4)  coordinate  (IItop)
       +(-90:4) coordinate (IIbot)
       +(0:4)   coordinate                  (IIright);
\draw     (IItop) -- (IIright) -- (IIbot) ;
\path 
   (I) +(90:4)  coordinate (Itop)
       +(-90:4) coordinate (Ibot)
       +(180:4) coordinate (Ileft);
\draw  (Itop) -- (Ileft) -- (Ibot); 
\draw[decoration={zigzag,pre=curveto,post=curveto},decorate] (IItop) to[bend right=15] node[midway, above, inner sep=2mm] {} (Itop);
\draw[decorate,decoration=zigzag] (IItop) -- (IIbot)
    node[midway, left, inner sep=2mm] {};
\draw[decorate,decoration=zigzag] (Itop) -- (Ibot);   
\draw[decoration={zigzag,pre=curveto,post=curveto},decorate] (IIbot) to[bend left=15] node[midway, below, inner sep=2mm] {} (Ibot)  ;
        \fill[blue] (-4,-1) circle (3.5pt);
        \fill[blue] (4,-1) circle (3.5pt);
        \draw[blue, thick] (-4,-1) -- (0,3.5);
        \draw[blue, thick] (4,-1) -- (0,3.5);
        \node (A) at (4.5,-1) {$t_c$};
\end{tikzpicture}
\caption{\la{penrose}The Penrose diagram of the D$p$-brane black hole, for $p<3$. Each point on the diagram corresponds to a spatial $S_{8-p}$. The $z$ coordinates \eqref{gzz} only cover the region I. The diagram is similar to the Penrose diagram of the AdS black brane. In addition to the black hole singularity (which bends down \cite{Fidkowski_2004}) there is a boundary singularity near $z=0$.
\label{fig:penrose}
}
\end{figure}

 In Lorentzian signature, the 2-pt function probes the quasinormal modes of the AdS$_{d+1}$ black brane metric. In momentum space it is fully determined by these quasinormal modes \cite{dodelson2023thermalproductformula}, which arise as poles in the complex $\omega$ plane. In the WKB approximation for large $\Delta$ (which is appropriate for the giant gravitons $\Delta \sim N$), the quasinormal modes are given by \cite{Festuccia_2009, biggs2023scaling}
\begin{align}\label{qnm}
    \alpha_n = \frac{2}{d}e^{-i\pi/d}\sqrt{\frac{d}{d-2}}\left(\frac{d-2}{d}\right)^{1/d}\left(\Delta-\frac{d}{2}+\left(n+\frac{1}{2}\right)\sqrt{d}\right),
\end{align}
where $n \geq 0$ is an integer.
 They are related to signatures of behind-the-horizon physics; we expect the exponential decay of the momentum space Wightman function $G_+(\omega)$ along the imaginary $\omega$ axis to be a signature of the black hole singularity \cite{Festuccia_2006}.  The rate of decay is given by 
 \begin{align}
     t_c=-\frac{\pi}{d}\cot\left(\frac{\pi}{d}\right)
 \end{align}
and is a measure of how much the singularity ``bends down", which is the phenomenon that leads to the presence of a ``bouncing geodesic" \cite{Fidkowski_2004}; see Figure \ref{fig:penrose}. In particular, this bouncing geodesic is a null geodesic with $E\to \infty$ that joins the points at $t=t_c$ on either boundary. $t$ here refers to the bulk time in terms of the coordinates in (\ref{gzz}). The time in the boundary theory $t'$ is related to $t$ via a rescaling $t'=\frac{(d_p(2 \pi)^{p-2} g_s N )^{1/(3-p)}}{\sqrt{\alpha'}}t$. Note also that the ratio of the real and imaginary parts of the quasinormal modes $\alpha_n$ in \eqref{qnm} is also proportional to $t_c$, and so dictates the slope of the quasinormal mode ``Christmas tree''.

\bibliography{main}
\bibliographystyle{JHEP}
\end{document}